\documentclass[aps,preprint,showkeys,showpacs]{revtex4}
\usepackage{tipa}
\usepackage{amssymb}
\usepackage{txfonts}
\usepackage{hyperref}
\usepackage{graphicx}
\usepackage{epsfig}
\begin{document}

\title{Parametrization of Tachyon Field }
\author{Ying Shao,\footnote{Email: sybb37@student.dlut.edu.cn}
Yuan-Xing Gui\footnote{Email: thphys@dlut.edu.cn}, Wei Wang}
\affiliation{Department of Physics, Dalian University of Technology,
Dalian 116023}

\begin{abstract}
We assume that universe is dominated by non-relativistic matter and
tachyon field and reconstruct the potential of tachyon field
directly from the effective equation of state (EOS) of dark energy.
We apply the method to four known parametrization of equation of
state and discuss the general features of the resulting potentials.
\end{abstract}
\keywords{parametrization; equation of state; reconstruction;
tachyon field.}

\pacs{98.80.Cq, 98.65.Dx}

\maketitle
\section{Introduction}
Observations of Type Ia supernovae indicate that our universe has
entered a phase of accelerated expansion in the recent past.$^{1,2}$
The accelerated expansion has been attributed to the existence of
mysterious dark energy with negative pressure. At present, there
have been many papers devoted to addressing the nature of the dark
energy. The cosmological constant $\Lambda$ is the simplest and most
obvious candidate for dark energy. However this model suffers from
the theoretical problems---fine-tuning and coincidence problems.
These problems have led to a variety of alternative models where the
dark energy component varies with time, such as
Quintessence,$^{3,4}$ Chaplygin gas,$^{5,6}$ modified gravity,$^{7}$
phantom,$^{8}$ K-essence$^{9}$ and so on. For the models of dark
energy with a scalar field (eg. quintessence, phantom, K-essence),
one can design many kinds of potentials and then study equation of
state (EOS) for the dark energy. On the other hand, the potential
can also be reconstructed from a parametrization of the EOS fitting
the observational data.$^{10}$ The latter has the advantage that it
does not depend on a specified model of dark energy and, therefore,
is also called a
model-independent method.$^{11}$ \\
\mbox{}\hspace{20pt}Recently it has been suggested that rolling
tachyon condensates, in a class of string theories, may have
interesting cosmological consequences. In this article, we
concentrate on the issue of the tachyon as a source of the dark
energy. The tachyon$^{12,13}$ is an unstable field which has become
important in string theory through its role in the Dirac-Born-Infeld
(DBI) action which is used to describe the D-brane action.$^{14-17}$
A number of authors have already demonstrated that the tachyon could
play a useful role in cosmology,$^{18-21}$ independent of the fact
that it is an unstable field. The tachyon can act as the source of
dark energy depending on the form of the associated
potential.$^{22-25}$ The purpose of this paper is to use the
model-independent method to reconstruct
the potential of tachyon field from the EOS of the dark energy.\\
\mbox{}\hspace{20pt}Various parametrization of the EOS of dark
energy has been presented and investigated. In this paper, we will
use four of them$^{26-30}$ to reconstruct the tachyon potential
$V(T)$ directly from the EOS $w_T(z)$ and then discuss the general
features of the resulting potentials. Moreover the difference
between tachyon and quintessence is also obtained for the evolution
of the potential. The outline of this paper is as follows: In
section II, we introduce the DBI action and the associated equation
of motion for the tachyon field. Furthermore, the potential of
tachyon field is reconstructed from EOS of dark energy $w_T(z)$.
Section III we apply the model independent method to the four
typical parametrizations of EOS and discuss the general features of
the resulting potentials. Section IV contains the conclusions.
\section{Reconstructing the Potential of Tachyon Field}
The Dirac-Born-Infeld (DBI) type effective 4-dimensional action is
described by$^{14,15}$
$$
S=\int
d^4x\bigg{[}\sqrt{-g}\frac{R}{2}-V(T)\times\sqrt{-det(g_{\mu\nu}+{\partial_\mu}T{\partial_\nu}T)}\bigg{]},\eqno(1)
$$
where $V(T)$ is the potential of tachyon field $T$. The above DBI
action is believed to describe the physics of tachyon condensation
for all values of $T$ as long as string coupling and
the second derivative of $T$ are small.\\
\mbox{}\hspace{20pt}We consider a cosmological scenario in which the
system is filled with non-relativistic matter and the tachyon field
$T$. In a flat FRW metric, the density $\rho_T$ and pressure $p_T$
of tachyon field are given by
$$
\rho_T=\frac{V(T)}{\sqrt{1-\dot{T}^2}},\eqno(2)
$$
$$
p_T=-V(T)\sqrt{1-\dot{T}^2}.\eqno(3)
$$
The EOS of the dark energy is
$$
w_T\equiv\frac{p_T}{\rho_T}=\dot{T}^2-1.\eqno(4)
$$
 The Friedmann equation can be written as:
$$
H^2=\frac{8\pi G}{3}(\rho_M+\rho_T),\eqno(5)
$$
where $\rho_M$ is density of non-relativistic matter.\\
\mbox{}\hspace{20pt}For a spatially homogeneous tachyon field, the
equation of motion is
$$
\dot{\rho_T}+3H(\rho_T+p_T)=0,\eqno(6)
$$
which yield
$$
\rho_T
=\rho_{T0}\exp\Big{[}3\int^z_0{(1+w_T)d\ln(1+z)}\Big{]},\eqno(7)
$$
where the dot denotes the differentiation with respect to $t$ and
subscript 0 represents the value of a quantity at present $(z=0)$.
The redshift $z$ is given by $1+z=a_0/a$. From Eqs.(2),(4) and (7),
we obtain
$$
V[T(z)]=\rho_{T0}\sqrt{-w_T}
\exp\Big{[}3\int^z_0{(1+w_T)d\ln(1+z)}\Big{]},\eqno(8)
$$
$$
\bigg{(}\frac{dT}{dz}\bigg{)}^2=\frac{1+w_T}{(1+z)^2H^2(z)},\eqno(9)
$$
where $dz/dt=-(1+z)H(z)$. With the help of
$\rho_M=\rho_{M_0}(1+z)^3$ and Eq.(7), the Friedmann equation (5)
becomes
$$
H(z)=H_0\bigg{[}\Omega_{M0}(1+z)^3+\Omega_{T0}\exp\Big{[}3\int^z_0{(1+w_T)d\ln(1+z)}\Big{]}\bigg{]}^{\frac{1}{2}},\eqno(10)
$$
where $\Omega_{M0}=\rho_{M0}/(3H_0^2/8\pi G),
\Omega_{T0}=\rho_{T0}/(3H_0^2/8\pi G)$ and
$\Omega_{M0}+\Omega_{T0}=1$. Substituting into Eq.(9), we have
$$
\frac{dT}{dz}=\pm\frac{\sqrt{1+w_T}}{H_0(1+z)\bigg{[}\Omega_{M0}(1+z)^3+(1-\Omega_{M0})\exp\Big{[}3\int^z_0{(1+w_T)d\ln(1+z)}\Big{]}\bigg{]}^{\frac{1}{2}}},\eqno(11)
$$
where the upper (lower) sign applies if $\dot{T}<0 (\dot{T}>0)$. As
the sign can be changed by the field redefinition, $T\rightarrow-T$,
it is arbitrary. Thus we choose the lower sign in the following
sections. Eqs.(8) and (11) are the potential and field function of
tachyon field which we have reconstructed.\\
\mbox{}\hspace{20pt}We define the following quantities
$$
\tilde{V}[T(z)]\equiv\frac{V[T(z)]}{\rho_{T0}},~~~~\tilde{T}(z)\equiv\frac{T(z)}{H^{-1}_0}.\eqno(12)
$$
Eqs.(8) and (11) can be written as
$$
\tilde{V}[T(z)]=\sqrt{-w_T}
\exp\Big{[}3\int^z_0{(1+w_T)d\ln(1+z)}\Big{]},\eqno(13)
$$
$$
\frac{d\tilde{T}}{dz}=\pm\frac{\sqrt{1+w_T}}{(1+z)\bigg{[}\Omega_{M0}(1+z)^3+(1-\Omega_{M0})\exp\Big{[}3\int^z_0{(1+w_T)d\ln(1+z)}\Big{]}\bigg{]}^{\frac{1}{2}}}.\eqno(14)
$$
\section{Parametrizations of the Potential $V(T)$}
Now, let us discuss the evolutions of potential $V(T)$ and field
function $T$ of tachyon field. It was shown in Ref.10 that the
quintessence potentials can be reconstructed from a given EOS of
dark energy $w_\phi(z)$. In this spirit, we can also reconstruct the
potential of tachyon field from a given concrete form of EOS
$w_T(z)$. The following four cases are considered: a constant EOS
parameter
and the other three two-parameter parametrizations.\\
\textbf{Case I:} $w_T=w_0$ (Ref.26). For this case, $w_T$ is a
constant and Eqs.(13) and (14) can be written by
$$
\tilde{V}(z)=\sqrt{-w_0}(1+z)^{3(1+w_0)},\eqno(15)
$$
$$
\frac{d\tilde{T}}{dz}=-\frac{\sqrt{1+w_0}}{(1+z)\bigg{[}\Omega_{M0}(1+z)^3+(1-\Omega_{M0})(1+z)^{3(1+w_0)}\bigg{]}^{\frac{1}{2}}}.\eqno(16)
$$
\textbf{Case II:} $w_T=w_0+w_1z$ (Ref.27). For this case, Eqs.(13)
and (14) can be given by
$$
\tilde{V}(z)=\sqrt{-w_0-w_1z}(1+z)^{3(1+w_0-w_1)}e^{3w_1z},\eqno(17)
$$
$$
\frac{d\tilde{T}}{dz}=-\frac{\sqrt{1+w_0+w_1z}}{(1+z)\bigg{[}\Omega_{M0}(1+z)^3+(1-\Omega_{M0})(1+z)^{3(1+w_0-w_1)}e^{3w_1z}\bigg{]}^{\frac{1}{2}}}.\eqno(18)
$$
\textbf{Case III:} $w_T=w_0+w_1\frac{z}{1+z}$ (Ref.28 and 29).
Similar to Cases I and II, for this case we have
$$
\tilde{V}(z)=\sqrt{-w_0-w_1\frac{z}{1+z}}(1+z)^{3(1+w_0+w_1)}e^{-3\frac{w_1z}{1+z}},\eqno(19)
$$
$$
\frac{d\tilde{T}}{dz}=-\frac{\sqrt{1+w_0+w_1\frac{z}{1+z}}}{(1+z)\bigg{[}\Omega_{M0}(1+z)^3+(1-\Omega_{M0})e^{-3\frac{w_1z}{1+z}}(1+z)^{3(1+w_0+w_1)}\bigg{]}^{\frac{1}{2}}}.\eqno(20)
$$
\textbf{Case IV:} $w_T=w_0+w_1\ln{(1+z)}$ (Ref.30). For this case,
Eqs.(13) and (14) can also be written
$$
\tilde{V}(z)=\sqrt{-w_0-w_1\ln(1+z)}(1+z)^{3(1+w_0)+\frac{3}{2}w_1\ln(1+z)},\eqno(21)
$$
$$
\frac{d\tilde{T}}{dz}=-\frac{\sqrt{1+w_0+w_1\ln(1+z)}}{(1+z)\bigg{[}\Omega_{M0}(1+z)^3+(1-\Omega_{M0})(1+z)^{3(1+w_0)+\frac{3}{2}w_1\ln(1+z)}\bigg{]}^{\frac{1}{2}}},\eqno(22)
$$
\begin{figure}
\includegraphics{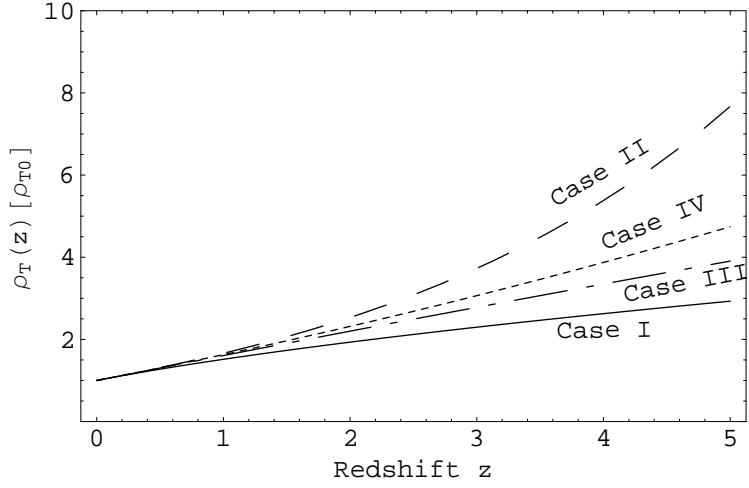}
\caption{Evolution of the energy density of tachyon $\rho_T(z)$.}
\label{Fig.1.}
\end{figure}
\begin{figure}
\includegraphics{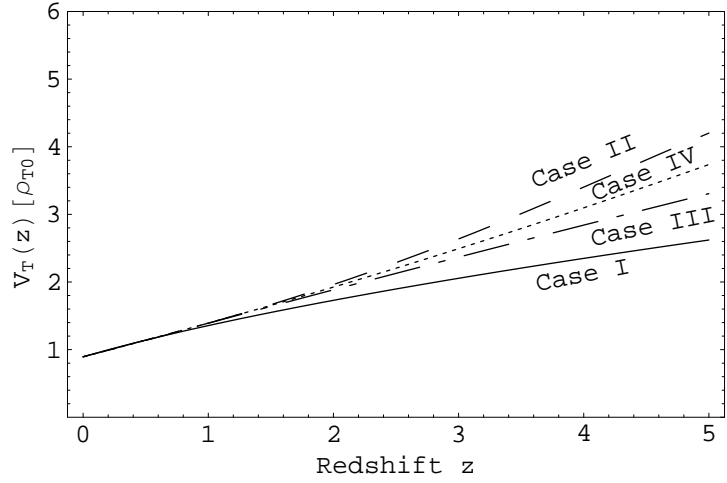}
\caption{Evolution of the potential of tachyon $V_T(z)$.}
\label{Fig.2.}
\end{figure}
\begin{figure}
\includegraphics{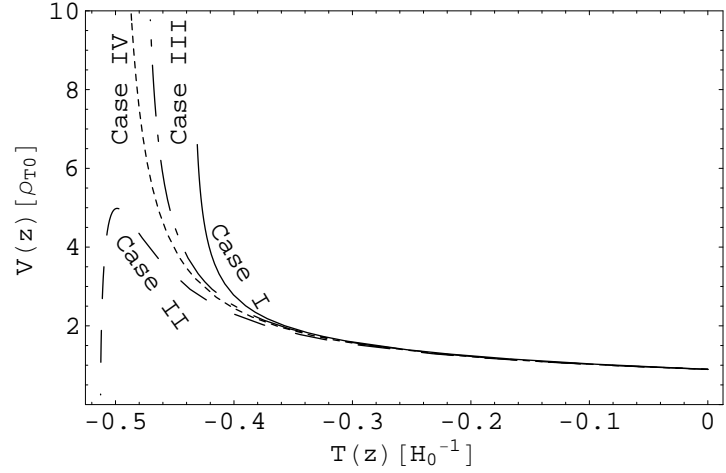}
\caption{Constructed tachyon potential $V(T)$.} \label{Fig.3.}
\end{figure}
\mbox{}\hspace{20pt}We have numerically evaluated the above
equations. For the specific reconstruction we choose $w_0=-0.8$,
$w_1=0.1$, $\Omega_{M0}=0.25$. Fig.1 and fig.2 show the evolutions
of the energy density $\rho_T(z)$ and potential $V_T(z)$ of the
tachyon with redshift $z$, respectively. At low redshift region, all
models share the same asymptotical behavior, but deviate from this
at redshift $z>1$. Furthermore, the relation of the reconstructed
potential $V(T)$ and field function $T$ is showed in fig.3. We see
that the four cases possess the same asymptotic behavior for the
region $T>-0.3$ and deviate from this at $T<-0.3$. Moreover from
fig.2 and 3, we note that when $z$ decreases, the potential $V(T)$
decreases and $T$ increases. This means that the potential $V(T)$
decreases as the universe expands. In addition we find there is
potential hill for the evolution of tachyon potential in Case II, which is different from other three cases.\\
\section{Conclusions}
In this paper, we have considered a spatially flat FRW universe
which is dominated by the non-relativistic matter and a spatially
homogeneous tachyon field $T$. By introducing Dirac-Born-Infeld
(DBI) action, we have obtained the equation of motion for tachyon
field. Then we have used the model-independent method to reconstruct
the potential of tachyon field from EOS of dark energy $w_T(z)$.
Furthermore, for four known EOS of dark energy we have plotted the
evolutions of the potential $V(T)$ and the energy density
$\rho_T(z)$. By analysis, we have given the following results:
\begin{itemize}
\item The potential $V_T(z)$ and energy density $\rho_T(z)$ for
the four cases share the same asymptotical behavior at low redshift
($0<z<1$) and deviate from this at redshift $z>1$. \item When EOS
parameter for dark energy $w_T(z)=w_0+w_1z$ (Case II), there is
potential hill for the evolution of tachyon potential, which is the
different from the other three cases. \item The reconstructed
tachyon potential $V(T)$ is in the form of a runaway potential at
$T>-0.5$.
\end{itemize}
\mbox{}\hspace{20pt}The above four parametrizations we have adopted
are directly related to the data of type Ia supernovae. Our approach
is a useful tool to reexamine the tachyon model. In the future, more
accurate observations could help greatly to determine the parameters
in the dark energy parametrization. By mapping of the recent
expansion history, we will learn more about the essence of the dark
energy.
\section*{ACKNOWLEDGMENTS}
This work was supported by National Science Foundation of China
under Grant NO.10573004.

\end{document}